\documentclass{ws-ijmpa}
\usepackage[super,compress]{cite}
\usepackage{graphicx}
\begin{document}
\markboth{G.~L.~Klimchitskaya}{Quantum field theory of the Casimir force for graphene}

\catchline{}{}{}{}{}
%
%%%%%%%%%%%%%%%%%%%%%%%%%%%%%%%%%%%%%%%%%%%%%%%%%%%%%%%%%%%%%%%%%%%%

\title{Quantum field theory of the Casimir force for graphene
}

\author{G.~L.~Klimchitskaya
}

\address{Department of Astrophysics,
Central Astronomical Observatory at Pulkovo of the Russian Academy of Sciences,
Saint Petersburg,
196140, Russia\\
and\\
Institute of Physics, Nanotechnology and
Telecommunications, Peter the Great Saint Petersburg
Polytechnic University, Saint Petersburg, 195251, Russia\\
g.klimchitskaya@gmail.com}

\maketitle

\begin{history}
\received{31 August 2015}
%\revised{Day Month Year}
\end{history}

\begin{abstract}
We present theoretical description of the Casimir interaction in graphene systems
which is based on the Lifshitz theory of dispersion forces and the formalism of
the polarization tensor in (2+1)-dimensional space-time. The representation for
the polarization tensor of graphene allowing the analytic continuation to the
whole plane of complex frequencies is given. This representation is used to
obtain simple asymptotic expressions for the reflection coefficients at all
Matsubara frequencies and to investigate the origin of large thermal
effect in the Casimir force for graphene. The developed theory is shown to
be in a good agreement with the experimental data on measuring the gradient
of the Casimir force between a Au-coated sphere and a graphene-coated substrate.
The possibility to observe the thermal effect for graphene due to a minor
modification of the already existing experimental setup is demonstrated.

\end{abstract}

\ccode{PACS numbers: 78.67.Wj, 12.20.Ds, 42.50.Nn, 68.65.Pq}

\section{Introduction}

Graphene is a two-dimensional sheet of carbon atoms possessing unusual properties,
which make it interesting for both fundamental physics and for numerous
applications.\cite{1} The electronic excitations in pristine (undoped) graphene
at frequencies below a few eV are massless and possess linear dispersion
relation,\cite{1,2} where the speed of light is replaced by the Fermi velocity
(which is the so-called Dirac model of graphene).
As a result, the interaction of the electromagnetic field with these excitations
is described by relativistic quantum electrodynamics in (2+1)-dimensions.
The existence of massless charged particles in graphene provides a unique
opportunity to test some fundamental predictions of quantum field theory which
are beyond the reach for usual elementary particles. Among the effects which have
become accessible, one could mention the Klein paradox in the interaction of
graphene with an electrostatic potential barrier\cite{3} and the creation of
graphene quasiparticles from vacuum either by the Schwinger mechanism in a
static electric field\cite{4,5} or in a time-dependent field.\cite{6}

One physical phenomenon, where the presence of graphene leads to outstanding results,
is the Casimir effect. It is well known that the Casimir force between closely spaced
material boundaries is caused by the zero-point and thermal fluctuations of the
electromagnetic field.\cite{7} If at least one of the boundary surfaces is formed by
a graphene sheet, the Casimir effect takes some new features. Specifically, it was
shown\cite{8} that for graphene the thermal correction to the Casimir force becomes
dominant at by an order of magnitude shorter separations than for bodies made of
ordinary materials. Although several formalisms have been used in the literature
to study the Casimir effect in different systems including
graphene,\cite{8,9,10,11,12,13,14,15,16,17,18}
the most straightforward approach, starting from the first principles of quantum
electrodynamics, is based on the Lifshitz theory of the Casimir force\cite{7} and
the polarization tensor of graphene in (2+1)-dimensions.
This tensor was explicitly calculated in Ref.~\citen{19} at zero temperature and
in the framework of thermal quantum field theory in Ref.~\citen{20} at any
Matsubara frequency. Currently the polarization tensor is actively used for
investigation of the Casimir effect in graphene systems and many important results
are obtained quite recently.\cite{21,22,23,24,25,26,27,28,29,30,31,32,33,34,35}

In this paper, we briefly describe the formalism of the polarization tensor and its
applications to graphene. In Sec.~2 we present  the Lifshitz formula for graphene
as an infinitesimally thin plane sheet and consider various forms of the reflection
coefficients. This allows one to make a link between different formalisms using the
spatially nonlocal dielectric permittivities of graphene, electric polarizabilities,
dynamic conductivities, and density-density correlation functions, on the one hand,
and the polarization tensor, on the other hand. Section~3 is devoted to different
representations of the polarization tensor. Here, main attention is given to the
recently discovered representation\cite{34} allowing an analytic continuation to the
entire plane of complex frequencies including the real frequency axis.
In Sec.~4, using this representation, the origin of the large thermal effect for
graphene is discussed. The measurements of the gradient of the Casimir force between
a Au-coated sphere and a graphene-coated substrate are considered in Sec.~5.
It is shown that the measurement data are in a very good agreement with theory using
the polarization tensor. Section~6 is devoted to the possibility of observing the large
thermal effect which is predicted for graphene systems at short separation
distances. We argue that this effect can be observed by means of already existing
experimental setup using the dynamic atomic force microscope (AFM). In Sec.~7
the reader will find our conclusions and discussion.

\section{Lifshitz Formula for Graphene}

The Lifshitz theory of van der Waals and Casimir forces was originally formulated for
the case of two semispaces described by the frequency-dependent dielectric
permittivities.\cite{7} At present, using the scattering approach, it is generalized
for bodies of arbitrary geometrical shape.\cite{36} It is only required that the
electromagnetic scattering amplitudes for each body be available.
For two plane parallel bodies separated by the vacuum gap of width $a$ at temperature
$T$ in thermal equilibrium the free energy per unit area takes common form, irrespective
of whether these bodies are material semispaces or infinitely thin graphene sheets:
\begin{eqnarray}
&&
{\cal F}(a,T)=\frac{k_BT}{2\pi}\sum_{l=0}^{\infty}
{\vphantom{\sum}}^{\prime}\int_{0}^{\infty}k_{\bot}\,dk_{\bot}
\left\{\ln\left[1-r_{{\rm TM}}^{(1)}(i\xi_l,k_{\bot})
r_{{\rm TM}}^{(2)}(i\xi_l,k_{\bot})
e^{-2aq_l}\right]\right.
\nonumber \\
&&~~~~~~~~~~~~~~~~~~~~~~~~~~~~~~~~~
+\left.\ln\left[1-r_{{\rm TE}}^{(1)}(i\xi_l,k_{\bot})
r_{{\rm TE}}^{(2)}(i\xi_l,k_{\bot})
e^{-2aq_l}\right]\right\}.
\label{eq1}
\end{eqnarray}
\noindent
Here, $r_{{\rm TM,\,TE}}^{(1)}$ and $r_{{\rm TM,\,TE}}^{(2)}$ are the reflection
coefficients for two independent
polarizations of the electromagnetic field, transverse magnetic (TM) and
transverse electric (TE), on the first  and second body, respectively.
The Matsubara frequencies are $\xi_l=2\pi k_BTl/\hbar$, where
$k_B$ is the Boltzmann constant,
$l=0,\,1,\,2,\,\ldots$, $q_l^2=k_{\bot}^2+\xi_l^2/c^2$,  where
$k_{\bot}=|\mbox{\boldmath$k$}_{\bot}|$ is the magnitude of the
projection of the wave vector on the plane of the
graphene sheet, and the prime on
the summation sign means that the term with $l=0$ is divided by two.
For ordinary material bodies $r_{{\rm TM,\,TE}}^{(1,2)}$ are the familiar
Fresnel reflection coefficients calculated along the imaginary frequency axis.
Below we consider the cases when at least one of the boundary bodies is a
graphene sheet either free-standing or deposited on a substrate. For a free-standing
graphene sheet
we notate the reflection coefficients as $r_{{\rm TM,\,TE}}^{(g)}$.

According to Sec.~1, the most fundamental quantity describing the response of
graphene on the electromagnetic field is the polarization tensor
$\Pi_{kn}=\Pi_{kn}(\omega,k_{\bot})$ with $k,\,n=0,\,1,\,2$.
This is a diagonal tensor, and only two of its components are independent.
It is customary to use $\Pi_{00}$ and
$\Pi_{\rm tr}=\Pi_0^{~0}+\Pi_1^{~1}+\Pi_2^{~2}$
as independent quantities.
In fact, the polarization tensor is closely related to other physical characteristics
of graphene. Keeping in mind applications to the Casimir effect, we present them at the
imaginary Matsubara frequencies. Thus, the longitudinal (along the surface) and
transverse electric susceptibilities (polarizabilities) of graphene
\begin{equation}
\alpha^{\|,\bot}(i\xi_l,k_{\bot})=
\varepsilon^{\|,\bot}(i\xi_l,k_{\bot})-1,
\label{eq2}
\end{equation}
\noindent
where $\varepsilon^{\|,\bot}(i\xi_l,k_{\bot})$ are the respective spatially nonlocal
dielectric permittivities, are expressed as\cite{30}
\begin{eqnarray}
&&
\alpha^{\|}(i\xi_l,k_{\bot})=\frac{1}{2\hbar k_{\bot}}
\Pi_{00}(i\xi_l,k_{\bot}),
\nonumber \\
&&
\alpha^{\bot}(i\xi_l,k_{\bot})=\frac{c^2}{2\hbar k_{\bot}\xi_l^2}
\Pi(i\xi_l,k_{\bot}).
\label{eq3}
\end{eqnarray}
\noindent
Here, the following notation is introduced:
\begin{equation}
\Pi(i\xi_l,k_{\bot})=k_{\bot}^2\Pi_{\rm tr}(i\xi_l,k_{\bot})-
q_l^2\Pi_{00}(i\xi_l,k_{\bot}).
\label{eq4}
\end{equation}

In a similar way, the longitudinal and transverse density-density correlation
functions of graphene are given by\cite{30}
\begin{eqnarray}
&&
\chi^{\|}(i\xi_l,k_{\bot})=-\frac{1}{4\pi\hbar e^2}
\Pi_{00}(i\xi_l,k_{\bot}),
\nonumber \\
&&
\chi^{\bot}(i\xi_l,k_{\bot})=-\frac{c^2}{4\pi\hbar e^2\xi_l^2}
\Pi(i\xi_l,k_{\bot}).
\label{eq5}
\end{eqnarray}
\noindent
Then, such important quantities as conductivities of graphene can be expressed
in the form\cite{17,30}
\begin{eqnarray}
&&
\sigma^{\|}(i\xi_l,k_{\bot})=-\frac{e^2\xi_l}{k_{\bot}^2}
\chi^{\|}(i\xi_l,k_{\bot})=\frac{\xi_l}{4\pi\hbar k_{\bot}^2}
\Pi_{00}(i\xi_l,k_{\bot}),
\nonumber \\
&&
\sigma^{\bot}(i\xi_l,k_{\bot})=-\frac{e^2\xi_l}{k_{\bot}^2}
\chi^{\bot}(i\xi_l,k_{\bot})=\frac{c^2}{4\pi\hbar k_{\bot}^2\xi_l}
\Pi(i\xi_l,k_{\bot}).
\label{eq6}
\end{eqnarray}

Equations (\ref{eq4}),  (\ref{eq5}) and (\ref{eq6}) make a link between the formalism of
the polarization tensor and other theoretical descriptions of graphene used in the literature.
In fact each of the formalisms can be used to express the reflection coefficients of graphene
$r_{{\rm TM,\,TE}}^{(g)}$. In so doing, the polarization tensor found in the one-loop
approximation turns out to be equivalent to the density-density correlation functions calculated
in the random-phase approximation. It should be taken into account, however, that until very recently
the full information about both longitudinal and transverse density-density correlation functions
(respectively, polarizabilities and conductivities) of graphene at nonzero temperature was not
available. Full information was obtained first for the polarization tensor of graphene at the
imaginary Matsubara frequencies.\cite{19,20}

The reflection coefficients on the free-standing graphene sheet expressed in terms of the
polarization tensor take the form\cite{19,20}
\begin{eqnarray}
&&
r_{{\rm TM}}^{(g)}(i\xi_l,k_{\bot})=
\frac{q_l\Pi_{00}(i\xi_l,k_{\bot})}{q_l\Pi_{00}(i\xi_l,k_{\bot})+2\hbar k_{\bot}^2},
\nonumber \\
&&
r_{{\rm TE}}^{(g)}(i\xi_l,k_{\bot})=-
\frac{\Pi(i\xi_l,k_{\bot})}{\Pi(i\xi_l,k_{\bot})+2\hbar k_{\bot}^2q_l},
\label{eq7}
\end{eqnarray}
\noindent
where $\Pi$ is defined in Eq.~(\ref{eq4}). Using Eqs.~(\ref{eq2}), (\ref{eq3}),
(\ref{eq5}) and (\ref{eq6}) the reflection coefficients can be expressed also in terms
of the polarizabilities of graphene, spatially nonlocal dielectric permittivities,
density-density correlation functions and conductivities.

\section{Two Representations for the Polarization Tensor}

The polarization tensor of graphene at zero temperature was first derived in Ref.~\citen{19}.
The generalization of the results of Ref.~\citen{19} for the case of nonzero temperature was
obtained in Ref.~\citen{20} at all imaginary Matsubara frequencies. It was extensively used
in investigations of the Casimir and Casimir-Polder forces in various
systems.\cite{22,23,24,25,26,27,28,29,30,31,32,33}

The polarization tensor of graphene at nonzero temperature can be presented as a sum of the
zero-temperature part and the thermal correction to it. For the 00-component and the quantity
defined in Eq.~(\ref{eq4}) we have
\begin{eqnarray}
&&
\Pi_{00}(i\xi_l,k_{\bot})=\Pi_{00}^{(0)}(i\xi_l,k_{\bot})+
\Delta_T\Pi_{00}(i\xi_l,k_{\bot}),
\nonumber \\
&&
\Pi(i\xi_l,k_{\bot})=\Pi^{(0)}(i\xi_l,k_{\bot})+
\Delta_T\Pi(i\xi_l,k_{\bot}),
\label{eq8}
\end{eqnarray}
\noindent
where $\Pi_{00}^{(0)}$ and $\Pi^{(0)}$ were derived at zero temperature with continuous
frequencies $\xi$, which were thereafter replaced with the Matsubara frequencies $\xi_l$.
The explicit expressions for $\Pi_{00}^{(0)}$ and $\Pi^{(0)}$ are well known.\cite{19}
For a graphene with nonzero mass gap parameter $\Delta$
(the electronic excitations in graphene can
acquire some small mass under the influence of electron-electron interactions, substrates
and defects of structure\cite{2}) they are given by
\begin{eqnarray}
&&
\Pi_{00}^{(0)}(i\xi_l,k_{\bot})=\frac{\alpha\hbar k_{\bot}^2}{\tilde{q}_l^2}
\Phi(\xi_l,k_{\bot}),
\nonumber \\
&&
\Pi^{(0)}(i\xi_l,k_{\bot})=\alpha\hbar k_{\bot}^2\Phi(\xi_l,k_{\bot}).
\label{eq9}
\end{eqnarray}
\noindent
Here, $\alpha$ is the fine-structure constant and the following notations are introduced:
\begin{eqnarray}
&&
\Phi(\xi_l,k_{\bot})=\frac{4}{\hbar}\left(c\Delta+
\frac{\hbar^2\tilde{q}_l^2-4 c^2\Delta^2}{2\hbar\tilde{q}_l}\,
\arctan\frac{\hbar\tilde{q}_l}{2c\Delta }\right),
\nonumber \\
&&
\tilde{q}_l^2=\frac{v_F^2}{c^2}k_{\bot}^2+\frac{\xi_l^2}{c^2}.
\label{eq9a}
\end{eqnarray}

The thermal correction to the polarization tensor is much more complicated.
In Ref.~\citen{20} it was obtained in the form which is valid only at the
imaginary Matsubara frequencies and does not allow analytic continuation to
the whole plane of complex frequencies including the real frequency axis.
Another representation for the thermal correction, which can be easily
analytically continued to the whole plane of complex frequencies, was derived
very recently.\cite{34} At the Matsubara frequencies, the thermal correction
of Ref.~\citen{34} takes the same values as that of Ref.~\citen{20}, but, as
opposed to the latter, satisfies all physical requirements at all other
frequencies. According to the results of Ref.~\citen{34}, the thermal
correction to the 00-component of the polarization tensor is given by
\begin{eqnarray}
&&
\Delta_T\Pi_{00}(i\xi_l,k_{\bot})=\frac{16\alpha\hbar c^2}{v_F^2}
\int_{0}^{\infty}\!\!dq_{\bot}\frac{q_{\bot}}{\Gamma(q_{\bot})}\,
\frac{1}{e^{\hbar c\Gamma(q_{\bot})/(k_BT)}+1}
\nonumber \\
&&~~~~~~~~~~~~
\times\left[1+\frac{1}{2}\sum_{\lambda=\pm 1}
\frac{M_{00,\,\lambda}(\xi_l,k_{\bot},q_{\bot})}{N_{\lambda}(\xi_l,k_{\bot},q_{\bot})}
\right].
\label{eq10}
\end{eqnarray}
\noindent
Here we use the notations:
\begin{eqnarray}
&&
N_{\lambda}(\xi_l,k_{\bot},q_{\bot})=\left[
Q_{\lambda}^2(\xi_l,k_{\bot},q_{\bot})-\left(
2\frac{v_F}{c}k_{\bot}q_{\bot}\right)^2\right]^{1/2},
\nonumber \\
&&
Q_{\lambda}(\xi_l,k_{\bot},q_{\bot})=\tilde{q}_l^2-2i\lambda\frac{\xi_l}{c}
\Gamma(q_{\bot}),
\nonumber \\
&&
M_{00,\lambda}(\xi_l,k_{\bot},q_{\bot})=4\Gamma^2(q_{\bot})-
\tilde{q}_l^2+4i\lambda\frac{\xi_l}{c}\Gamma(q_{\bot}),
\nonumber \\
&&
\Gamma(q_{\bot})=\sqrt{q_{\bot}^2+\frac{c^2\Delta^2}{\hbar^2}}.
\label{eq11}
\end{eqnarray}
\noindent
The thermal correction to the quantity (\ref{eq4}) takes the form\cite{34}
\begin{eqnarray}
&&
\Delta_T\Pi(i\xi_l,k_{\bot})=\frac{16\alpha\hbar c^2}{v_F^2}
\int_{0}^{\infty}\!\!dq_{\bot}\frac{q_{\bot}}{\Gamma(q_{\bot})}\,
\frac{1}{e^{\hbar c\Gamma(q_{\bot})/(k_BT)}+1}
\nonumber \\
&&~~~~~~
\times\left[-\frac{\xi_l^2}{c^2}+\frac{1}{2}\sum_{\lambda=\pm 1}
\frac{M_{\lambda}(\xi_l,k_{\bot},q_{\bot})}{N_{\lambda}(\xi_l,k_{\bot},q_{\bot})}
\right],
\label{eq12}
\end{eqnarray}
\noindent
where
\begin{equation}
M_{\lambda}(\xi_l,k_{\bot},q_{\bot})=\frac{\xi_l^2}{c^2}\tilde{q}_l^2-
4\Gamma^2(q_{\bot})\tilde{q}_l^2+
4\frac{v_F^2\Delta^2}{\hbar^2}k_{\bot}^2
-4i\lambda\frac{\xi_l}{c}\Gamma(q_{\bot})\tilde{q}_l^2.
\label{12a}
\end{equation}

%%%%%%__Figure__1%%%%%%%%%%%%%%%%%%%%%%%%%
\begin{figure}[b]
\vspace*{-7.5cm}
\hspace*{1.cm}
\centerline{\includegraphics[width=14cm]{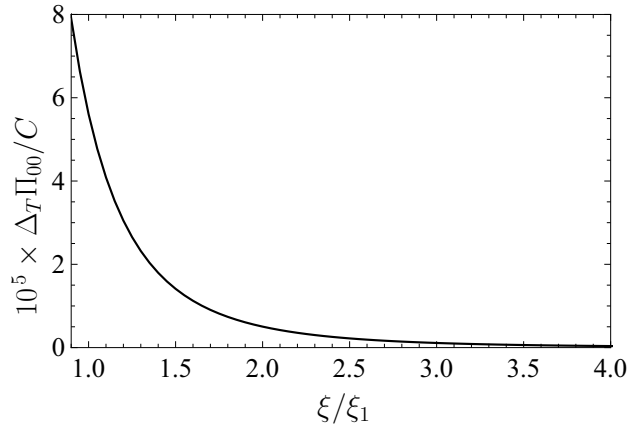}}
\vspace*{-7.cm}
\caption{The normalized thermal correction to the 00-component
of the polarization tensor of graphene computed as a function of the normalized
imaginary frequency at $k_{\bot}=10\xi_1$ using the representation of
Ref.~\citen{34}.}
\end{figure}
%%%%%%%%%%%%%%%%%%%%%%%%%%%%%%%%%%%%%%%%%%%%%%%%%%%%%%%%%%%%%%%%%%%%%
As an illustration, in Fig.~1 we plot the thermal correction
$\Delta_T\Pi_{00}(i\xi,k_{\bot})/C$ from Eq.~(\ref{eq10}), where the normalization
factor is $C=16\alpha ck_BT/v_F^2$ and the  frequency $\xi$ varies continuously
along the imaginary frequency axis. The thermal correction is shown as a function of
dimensionless variable $\xi/\xi_1$. The computations were performed for the gapless
graphene ($\Delta=0$) at $T=300\,$K and $k_{\bot}=10\xi_1$.
As is seen in Fig.~1, the thermal correction is a monotonously decreasing function of
the frequency. Similar results hold at any temperature and transverse wave vector.

%%%%%%__Figure__2_%%%%%%%%%%%%%%%%%%%%%%%%%
\begin{figure}[t]
\vspace*{-7.5cm}
\hspace*{1.cm}
\centerline{\includegraphics[width=14cm]{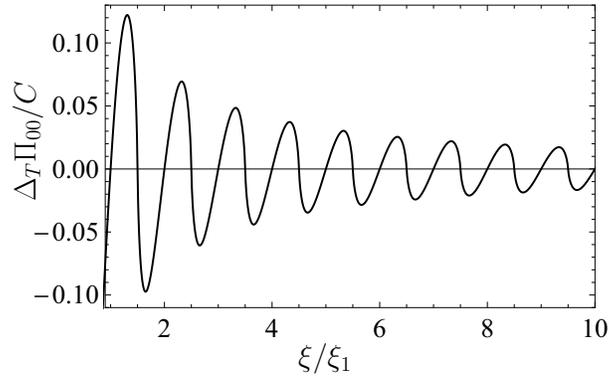}}
\vspace*{-7.cm}
\caption{The normalized thermal correction to the 00-component
of the polarization tensor of graphene computed as a function of the normalized
imaginary frequency at $k_{\bot}=10\xi_1$ using the representation of
Ref.~\citen{20}.}
\end{figure}
%%%%%%%%%%%%%%%%%%%%%%%%%%%%%%%%%%%%%%%%%%%%%%%%%%%%%%%%%%%%%%%%%%%%%
For comparison purposes, in Fig.~2 we present the normalized thermal correction
$\Delta_T\Pi_{00}(i\xi,k_{\bot})/C$ as a function of $\xi/\xi_1$, where the polarization
tensor $\Pi_{kn}$ is taken from Ref.~\citen{20}. The same values of $\Delta$, $T$ and
$k_{\bot}$, as in Fig.~1, are used in computations. As is seen in Fig.~2,
the function $\Delta_T\Pi_{00}(i\xi,k_{\bot})/C$ of Ref.~\citen{20} oscillates with the
period equal to $2\pi/\xi_1$ and decreasing amplitude. At the points
$\xi\equiv\xi_l=l\xi_1$ it takes the same values as
the function $\Delta_T\Pi_{00}(i\xi,k_{\bot})/C$ of Ref.~\citen{34} (see Fig.~1),
which are much smaller than the oscillation amplitudes. As a result, the analytic
continuation of the function $\Delta_T\Pi_{00}(i\xi,k_{\bot})$ of Ref.~\citen{20}
to the whole plane of complex frequencies becomes nonphysical.\cite{34}

\section{Origin of Large Thermal Effect for Graphene}

The polarization tensor of graphene in combination with Eqs.~(\ref{eq1}) and (\ref{eq7})
was used in many papers devoted to investigation of the Casimir effect in layered systems
including graphene sheets. Specifically, the Casimir force between a graphene sheet and
a material plate made of real dielectric or metal was investigated,\cite{22} as well as
between two graphene sheets.\cite{24} The polarization tensor was also used to explore
the Casimir-Polder interaction between different atoms and a graphene sheet\cite{23,26,27}
and the Casimir force between thin films\cite{25} and graphene-coated substrates.\cite{31}

The present section is devoted to the origin of thermal effect in the Casimir force for graphene
which becomes dominant at much shorter separations than for ordinary materials. This origin can be
investigated in the most straightforward way using the representation (\ref{eq10}) and
(\ref{eq12}) for the polarization tensor. The point is that at all $l\geq 1$ and $T=300\,$K
the expressions (\ref{eq10}) and (\ref{eq12}) are considerably simplified.
Taking into account that $\xi_1\approx 2.4\times 10^{14}\,$rad/s and that the characteristic
photon wave number, giving the major contribution to the Casimir free energy and pressure,
is $k_{\bot}=1/(2a)$, one finds that at $a\geq 10\,$nm there is the natural small parameter
\begin{equation}
\frac{4v_F^2k_{\bot}^2}{c^2\tilde{q}_l^2}<
\frac{4v_F^2k_{\bot}^2}{c^2\tilde{q}_1^2}\ll 1.
\label{eq13}
\end{equation}

Expanding Eqs.~(\ref{eq10}) and (\ref{eq12}) for the case $\Delta=0$ in powers of the small
parameter (\ref{eq13}), one obtains\cite{35}
\begin{eqnarray}
&&
\Delta_T\Pi_{00}(i\xi_l,k_{\bot})=\frac{\alpha\hbar k_{\bot}^2}{\tilde{q}_l}\,Y_l,
\nonumber \\
&&
\Delta_T\Pi(i\xi_l,k_{\bot})={\alpha\hbar k_{\bot}^2}{\tilde{q}_l}Y_l,
\label{eq14}
\end{eqnarray}
\noindent
where
\begin{equation}
Y_l\equiv4\int_{0}^{\infty}\frac{du}{e^{\pi lu}+1}\,\frac{u^2}{1+u^2}.
\label{eq15}
\end{equation}
\noindent
Using Eq.~(\ref{eq9}) for $\Delta=0$ and Eqs.~(\ref{eq14}), (\ref{eq8}) and (\ref{eq7}),
one arrives at
\begin{eqnarray}
&&
r_{\rm TM}(i\xi_l,k_{\bot})=
\frac{\alpha q_l(\pi +Y_l)}{\alpha q_l(\pi +Y_l)+2\tilde{q}_l},
\nonumber \\[1mm]
&&
r_{\rm TE}(i\xi_l,k_{\bot})=-
\frac{\alpha \tilde{q}_l(\pi +Y_l)}{\alpha\tilde{q}_l(\pi +Y_l)+2{q}_l}.
\label{eq16}
\end{eqnarray}
\noindent
Note that at $l=0$ the expressions (\ref{eq10}) and (\ref{eq12}) simplify considerably and their exact form is used in computations.\cite{35}

We illustrate the precision of approximate expressions (\ref{eq16}) in calculation of the
Casimir pressure between two graphene sheets,
\begin{equation}
P(a,T)=-\frac{\partial{\cal F}(a,T)}{\partial a},
\label{eq17}
\end{equation}
\noindent
where ${\cal F}(a,T)$ is given in Eq.~(\ref{eq1}). In Fig.~3 we plot the quantity
\begin{equation}
\delta{P}^{(k)}(a,T)=\frac{{P}(a,T)-{P}^{(k)}(a,T)}{{P}(a,T)}
\label{eq18}
\end{equation}
\noindent
at $T=300\,$K, $\Delta=0$ as a function of separation, where $P(a,T)$ is computed
exactly by Eqs.~(\ref{eq1}), (\ref{eq17}), (\ref{eq7})--(\ref{eq12}),
$P^{(1)}(a,T)$ is computed using the approximate expressions (\ref{eq16}) at $l\geq 1$
and $P^{(2)}(a,T)$ is computed by omitting the thermal correction to the
polarization tensor at all $l\geq 1$, but replacing the continuous frequencies with
the Matsubara ones\cite{20} (the lines 1 and 2, respectively). As is seen in Fig.~3,
the error arising from the use of approximate expressions (\ref{eq16}) is negative and
the maximum of its magnitude is equal to only 0.18\%. At $a\geq 35\,$nm our calculation
method\cite{35} becomes practically exact. The error in the second approximate
method\cite{20} is positive and its maximum value is equal to 0.87\%.
%%%%%%__Figure__3___%%%%%%%%%%%%%%%%%%%%%%%%%
\begin{figure}[t]
\vspace*{-8.5cm}
\hspace*{1.cm}
\centerline{\includegraphics[width=14cm]{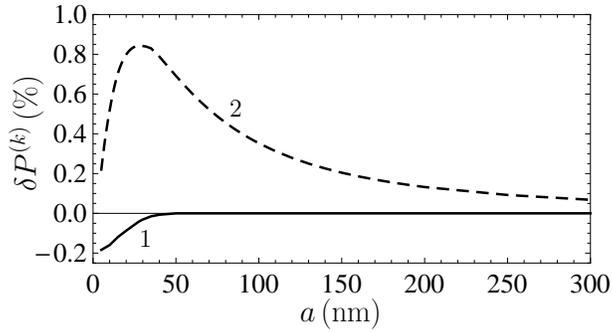}}
\vspace*{-7.cm}
\caption{The relative errors of the approximate methods for calculation
of the Casimir pressure between two graphene sheets using the asymptotic
approach accounting for an explicit temperature dependence of the polarization
tensor in all Matsubara terms (line 1) and disregarding this dependence
in terms with $l\geq 1$ (line 2) are shown as functions of separation.}
\end{figure}
%%%%%%%%%%%%%%%%%%%%%%%%%%%%%%%%%%%%%%%%%%%%%%%%%%%%%%%%%%%%%%%%%%%%%

Now we are in a position to calculate different contributions to the thermal Casimir
pressure between two graphene sheets with $\Delta=0$.
%%%%%%__Figure__4___%%%%%%%%%%%%%%%%%%%%%%%%%
\begin{figure}[b]
\vspace*{-7.7cm}
\hspace*{1.cm}
\centerline{\includegraphics[width=14cm]{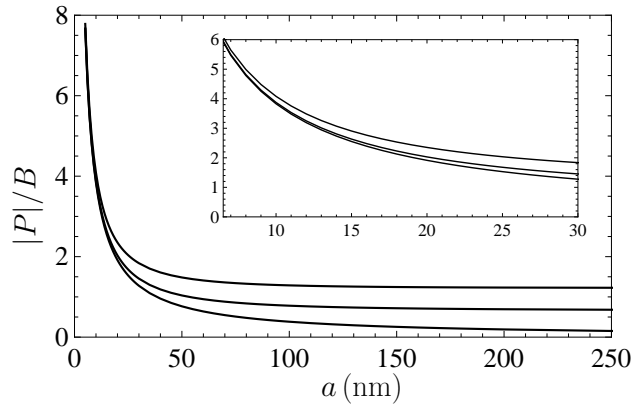}}
\vspace*{-7.cm}
\caption{The normalized Casimir pressure between two graphene sheets at $T=300\,$K
calculated at zero temperature (the lowest line), taking into account only an implicit
temperature dependence (the intermediate line) and exactly (the upper line) are shown as
functions of separation. The inset shows the region of short separations on an enlarged
scale.}
\end{figure}
%%%%%%%%%%%%%%%%%%%%%%%%%%%%%%%%%%%%%%%%%%%%%%%%%%%%%%%%%%%%%%%%%%%%%
In Fig.~4 we plot the magnitudes of the Casimir pressure normalized to the quantity
$B=k_BT/(8\pi a^3)$ at $T=300\,$K as functions of separation. The lowest, intermediate
and upper lines were computed at $T=0\,$K using the polarization tensor (\ref{eq9})
with the continuous frequency $\xi$,
at $T=300\,$K using the polarization tensor (\ref{eq9})
with the discrete Matsubara frequencies, and
at $T=300\,$K using the polarization tensor (\ref{eq10}) and (\ref{eq12}), respectively.
In an inset the region of short separations is shown on an enlarged scale.
Note that the same results for the upper line in Fig.~4 are obtained by using the approximate
reflection coefficients given in Eq.~(\ref{eq16}). The total thermal correction is
characterized by the difference of the upper and lowest lines in Fig.~4.
It consists of the explicit thermal effect originating from the dependence of the
polarization tensor on the temperature as a parameter (the difference between the upper and
intermediate lines) and the implicit thermal effect originating from a summation over the
temperature-dependent Matsubara frequencies (the difference between the intermediate and
lowest lines).

As is seen in Fig.~4, the thermal effect quickly increases with increasing separation.
At $a>150\,$nm it contributes more than 80\% of the Casimir pressure. In so doing, the
explicit thermal effect contributes more than the implicit one at the separations below
325\,nm. At larger separations the implicit thermal effect dominates over the explicit
one (this separation region is not shown in the figure). Thus, both contributions to
the thermal correction in the Casimir pressure between two graphene sheets are important
and should be taken into account.

 \section{Measuring of the Casimir Force for Graphene}

The first experiment on measuring the gradient of the Casimir force between a Au-coated
sphere and a graphene-coated substrate was performed\cite{37} by means of a dynamic atomic
force microscope (see Refs.~\citen{7} and \citen{38} for a description of this measurement
technique in application to the Casimir interaction). The sphere radius was $R=54.1\,\mu$m,
and the graphene sheet was deposited on a 300\,nm thick SiO${}_2$ film covering a B-doped
Si plate of $500\,\mu$m thickness.\cite{37}

The gradient of the Casimir force in sphere-plate geometry can be calculated using the
proximity force approximation as
\begin{equation}
\frac{\partial F_{sp}(a,T)}{\partial a}=2\pi R
\frac{\partial{\cal F}(a,T)}{\partial a},
\label{eq19}
\end{equation}
\noindent
where the Casimir free energy per unit area in the configuration of two parallel plates is
defined in Eq.~(\ref{eq1}). In this case the reflection coefficients $r_{\rm TM,\,TE}^{(1)}$
refer to a Au surface, whereas $r_{\rm TM,\,TE}^{(2)}$ describe the reflection properties
of graphene deposited on a two-layer substrate. Both the reflectivity of Au and of a
two-layer substrate can be easily expressed in terms of the frequency-dependent dielectric
permittivities of Au, SiO${}_2$ and Si. There was a problem, however, how to calculate the
combined reflection coefficient of a two-dimensional graphene sheet, deposited on a
three-dimensional substrate, where graphene is described by the polarization tensor and
substrate materials by the dielectric permittivity. This problem was solved in Ref.~\citen{31}
by the method of multiple reflections. The computations were performed using the polarization
tensor with the nonzero mass gap parameter $\Delta\leq 0.1\,$eV taking into account that the
graphene sheet could have the defects of structure and was deposited on a substrate.

The computed gradients of the Casimir force have been compared with the experimental data
obtained in two series of measurements made for two graphene-coated plates over the separation
region from 224 to 500\,nm. A very good agreement between the measurement data and theory
using the polarization tensor was demonstrated in the limit of experimental errors and
theoretical uncertainties for both graphene samples. The experimental data was shown to be
in complete disagreement with theoretical force gradients computed for the same substrate
with no graphene coating.\cite{37} The same measurement data were compared\cite{33} with the
alternative theoretical predictions obtained in the framework of the hydrodynamic model of
graphene.\cite{6,7,39} It was shown\cite{33} that the hydrodynamic model is excluded by the
measurement data at a 99\% confidence level over the wide region of separations.
Thus, the experiment of Ref.~\citen{37} can be considered as a confirmation of the Dirac
model of graphene.

\section{Possibility to Observe the Thermal Effect}

The experiment\cite{37} described in previous section has clearly demonstrated the influence
of graphene coating on the Casimir interaction. In Ref.~\citen{32} the problem was raised
on whether or not the same experimental data can be used as an evidence for large thermal
effect in the Casimir interaction predicted for graphene at short separations.\cite{8,22,24}
To solve this problem, the gradients of the Casimir force $F_{sp}^{\prime}$ in the
experimental configuration were computed as described above at $T=300\,$K and at $T=0\,$K
taking into account all theoretical uncertainties.\cite{32} It was shown that within the
separation region from 224 to 300\,nm the obtained theoretical bands do not overlap.
The experimental data with their errors presented as crosses were shown to mostly belong
to the theoretical band computed at $T=300\,$K (a few $\mu$N/m larger force gradients than
at $T=0\,$K). Only several data crosses slightly touch the theoretical band at $T=0\,$K,
which represents the force gradients with no thermal correction. This means that the
already performed experiment\cite{37} was only one step away from the conclusive
observation of large thermal effect in the Casimir interaction between a Au-coated sphere
and a graphene-coated substrate.

To solve a question as to whether the thermal effect is observable at the cost of a minor
modification of the existing laboratory setup, the Casimir pressure between two parallel
plates made of different materials was calculated, where at least one of the plates was
coated with graphene.\cite{32} Note that due to Eqs.~(\ref{eq17}) and  (\ref{eq19}),
calculations of the Casimir pressure are equivalent to the calculations of the force
gradients in the sphere-plate geometry, which is of an experimental interest.

The computations of the Casimir pressure were performed for the plates made of Au, Si,
sapphire, mica and SiO${}_2$. It was shown\cite{32} that the influence of graphene coating
(and of the thermal effect)is the most pronounced for a plate material having the smallest
static dielectric permittivity (SiO${}_2$).  It is a happy accident that in the experiment
of Ref.~\citen{37} just the amorphous SiO${}_2$ (silica) satisfying this condition has been
used as a material of the film underlying graphene. Then calculations of the thermal
correction to the gradient of the Casimir force between a Au-coated sphere and a graphene-coated
SiO${}_2$ plate at the experimental separations have been performed.
It was shown\cite{32} that within
the separation region from 224 to 350\,nm the thermal correction markedly (up to a factor of
five) exceeds the total experimental error. This makes the thermal effect observable if the
thickness of a SiO${}_2$ film is sufficiently large (computations were performed for a thick
SiO${}_2$ plate equivalent to a semispace).

Thus, the single point to be modified in the experiment of Ref.~\refcite{37} is the thickness
of a SiO${}_2$ film which was equal to only 300\,nm (see Sec.~5). If it were increased up to
at least $2\,\mu$m (providing almost the same effect as a SiO${}_2$ semispace) the large
thermal effect in the gradient of the Casimir force originating from graphene would be
observed with certainty.

\section{Conclusions and Discussion}

{}From the foregoing one can conclude that the Lifshitz theory and the formalism of the
polarization tensor in (2+1)-dimensional space-time provide fundamental field-theoretical
description of the Casimir interaction in layered systems containing graphene.
The representation for the polarization tensor allowing the analytic continuation to the
whole plane of complex frequencies can be also used to describe other physical phenomena,
e.g., the reflectivities of graphene and graphene-coated substrates.\cite{34,40}
This representation also leads to simple asymptotic expressions for the reflection
coefficients at all nonzero Matsubara frequencies which result in nearly exact values
for the Casimir energy and pressure. The developed theory is found to be in a very good agreement
with the measurement data of the first experiment on measuring the gradient of the Casimir
force between a Au-coated sphere and a graphene-coated substrate. In the framework of this
theory the large thermal effect predicted for graphene earlier was shown to consist of two
equally important parts originating from the parametric temperature dependence of the
polarization tensor and from a summation over the Matsubara frequencies. It is shown that
the thermal effect can be observed with already existing experimental setup if to make
a minor modification of the used graphene sample.

In the future it is desirable to generalize the developed formalism for the case of
nonzero chemical potential. This problem was recently solved,\cite{41} but only for the
case of gapless graphene.

\section*{Acknowledgments}
The author is grateful to M.~Bordag and V.~M.~Mostepanenko
for helpful discussions.

\end{document}